\documentclass[reprint, amsmath, amssymb, aps, prb, longbibliography, floatfix]{revtex4-1}
\usepackage{graphicx}% Include figure files
\usepackage{color}
\usepackage{bm}% bold math 
\usepackage[breaklinks=true,colorlinks,citecolor=blue,linkcolor=blue,urlcolor=blue]{hyperref}
\usepackage{float}
\makeatletter
\newcommand{\printfnsymbol}[1]{%
  \textsuperscript{\@fnsymbol{#1}}%
}

\def \be {\begin{equation}}
\def \ee {\end{equation}}
\def \bea{\begin{eqnarray}}
\def \eea{\end{eqnarray}}
\def \nn{\nonumber}

\begin{document}
%\selectlanguage{english}

%\title{Nonlinear magnetotransport in Weyl semimetal in weak magnetic field} % without tilt 
\title{Nonlinear magnetotransport in Weyl semimetal}
\author{Debottam Mandal}
\email{dmandal@iitk.ac.in}
\affiliation{Department of Physics, Indian Institute of Technology Kanpur, Kanpur 208016}
\author{Kamal Das}
\email{kamaldas@iitk.ac.in}
\affiliation{Department of Physics, Indian Institute of Technology Kanpur, Kanpur 208016}
\author{Amit Agarwal}
\email{amitag@iitk.ac.in}
\affiliation{Department of Physics, Indian Institute of Technology Kanpur, Kanpur 208016}

\begin{abstract}
The recent discovery of the quantum nonlinear Hall effect has revived the field of nonlinear transport. Here, we predict magnetic field induced nonlinear Hall effect in time reversal symmetric Weyl semimetal. We show that the interplay of the band geometric quantities, such as the Berry curvature, and the magnetic part of the Lorentz force can give rise to finite nonlinear Hall conductivity that is linear in the magnetic field. Such nonlinear Hall conductivity can manifest through nonlinear transport measurement as well as nonlinear optical phenomena like photocurrent and the second harmonic generation.
\end{abstract}

\maketitle
\section{Introduction}
%
%{\bf background of Weyl semimetals}

Weyl semimetals (WSM) are well known for hosting low energy quasi-particle excitation which mimic the properties of Weyl fermions~[\onlinecite{Wan2011,Weng2015,Lv2015,Xu2015,Xu2015b,Hasan2017,Yan2017,Amritage2018}]. Their novel bulk electronic structure comprises of doubly degenerate linear band crossing of non-degenerate bands, known as Weyl nodes. In addition, these materials support non-trivial and exotic  Fermi arc states on their surface. Intense research focused on the impact of  Weyl quasi-particles on physical  properties has led to the discovery of several novel bulk phenomena~[\onlinecite{lv_RMP2021_expt}] such as the quantum anomalies~[\onlinecite{chernodub_arxiv2021_thermal}], and some unconventional surface phenomena~[\onlinecite{Nishihaya2019,Nishihaya2021}]. Many of the bulk properties of WSM can be understood in terms of the Berry curvature associated with Weyl nodes, which act as source and sink of Berry curvature depending on the chirality of the nodes. 
The Weyl semimetal phase with space inversion symmetry (SIS) is realized in some magnetic systems ~[\onlinecite{Wan2011,Sagayama2013,Disseler2014}], and it can also be induced in Dirac semimetals such as Cd$_3$As$_{2}$~[\onlinecite{Wang2013}], and Na$_3$Bi~[\onlinecite{Wang2012}] by a magnetic field. Weyl systems with time reversal symmetry (TRS) have been realized in the TaAs family~[\onlinecite{Weng2015,Lv2015,SMHuang2015,Xu2015}], amongst others. In addition to these, WSM where both the symmetries are broken~[\onlinecite{Zyuzin2012}] have been recently realized in $R$AlGe family ($R=$ rare-earth)~[\onlinecite{Chang2018}] and in CeAlSi~[\onlinecite{yang_PRB2021_non}].

%{\bf non-linear response in Weyl semimetal}

The realization of WSM in SIS broken materials has further promoted the exploration of second order nonlinear (NL) responses~[\onlinecite{Boyd2008,Orenstein2021,SKDas2021}] in them. It has been shown that due to its topological aspects, the photogalvanic responses~[\onlinecite{Chan17, deJuan17,Golub17,Kharzeev18,Golub18,Parker2019,Sadhukhan2021a,Sadhukhan2021b}] and the second harmonic generation~[\onlinecite{Li2020}] show novel behaviour in WSM. Furthermore, the recently discovered Berry curvature dipole induced NL anomalous Hall effect~[\onlinecite{Deyo2009,Moore2010,Sodemann2015,QMa2019,Du2021,Ortix2021}] has also been realized in   WSM~[\onlinecite{Konig17,Rostami2018,Zeng2021}] in the absence of any magnetic field. Owing to these, it is expected that the NL responses of WSM in the presence of a magnetic field will also incorporate rich physics~[\onlinecite{Cortijo2016,Morimoto2016c,Nandy2021}]. In the strong magnetic field limit, the NL conductivities have been found to show quantum oscillation behavior~[\onlinecite{Golub17,Golub18}] owing to the presence of Landau levels, while the Berry curvature induced corrections in the semiclassical equation of motion~[\onlinecite{Gao2019}], has been shown to give rise to non-trivial responses in the weak magnetic field regime ~[\onlinecite{Cortijo2016, Morimoto2016c,gorbar_PRB2017_sec,Golub20,Zyuzin2018,Zyuzin18,Li2021}].  

%{\bf what we have done in this paper}

 Here, we predict a new magnetic field induced NL Hall effect in WSM with time reversal symmetry.  Using the semiclassical Boltzmann transport formalism in the weak magnetic field regime, we show that the SIS broken WSM possesses NL Hall conductivities that vary linearly with the magnetic field. These arise from  the interplay between the band geometric quantities, and the magnetic field component of the Lorentz force.  

More specifically, we show that 
the NL Hall conductivity with the same last two indices can be expressed as $\sigma_{abb} \propto B_a \delta_{{\hat {\bm a}} \times {\hat {\bm b}}, {\hat {\bm c}}}$, where $a,b,c$ are the orthogonal Cartesian coordinates. Such NL Hall conductivity (perpendicular to the applied electric field) drives current along the direction of the magnetic field. The physical mechanisms responsible for this are the Lorentz force, the Berry curvature dependent correction to the phase-space factor and the Berry curvature induced magnetic velocity correction. Furthermore, the NL Hall conductivity with the different last two indices can be expressed as $\sigma_{aab} \propto B_b$. For such NL conductivities current flows perpendicular to the direction of the magnetic field. The mechanism behind this contribution are the Lorentz force, correction to the phase-space factor and the `Berry force'. These novel NL Hall conductivity can be measured through NL resistivity measurements in magneto-transport experiments~[\onlinecite{He2019,Pacchioni2019,Kang2019,Shvetsov2019}]. In addition they can also manifest through nonlinear optical experiment of photocurrent and second harmonic generation.

The rest of the paper is organised as follows: In Sec.~\ref{BTF} we discuss the semiclassical Boltzmann transport formalism and derive the generic forms of the NL conductivities. We present our model specific calculations of the NL conductivities for the inversion symmetry broken WSM in Sec.~\ref{model_WM}. 
%This is followed by a discussions in Sec.~\ref{discussion} and 
Finally we summarize our results in Sec.~\ref{conclusion}.

\section{Semiclassical theory for nonlinear conductivities}
\label{BTF}

In this section, we present the general expressions of second order NL conductivities in quantum materials in presence of a magnetic field. For an AC electric field, two different NL conductivities are commonly defined. The second harmonic conductivity relates the applied electric field to the second harmonic current [${\bm j}^{(2\omega)}_2$] via the phenomenological relation $j^{(2\omega)}_{2,a} = \sigma_{abc} E_b E_c$. Here, sum over the repeated spatial indices ($a,b,c$) is implied. The DC or rectification NL conductivity ($\sigma_{abc}^R$) relates the rectification current [${\bm j}^{(0)}_2$] to the applied electric field through $j^{(0)}_{2,a} = \sigma_{abc}^R E_b E_c^*$. In this work we primarily focus on the second harmonic conductivity. 
%, and comment briefly on the second one.

To calculate the NL current, we employ the semiclassical Boltzmann transport formalism. In this formalism, the electrical current is expressed as ${\bm j}= -e \int [d{\bm k}] D^{-1} \dot {\bm r} g(t)$, where `$-e$' is the electronic charge and $[d{\bm k}]=d{\bm k}/(2\pi)^3$. It is evident from the above expression that we need three key ingredients for calculating the  current. These are $i)$ the equations of motion of the carriers, $ii)$ the phase-space density $D^{-1}$ and, $iii)$ the non-equilibrium distribution function (NDF) $g(t)$.

In presence of a homogeneous time dependent electric field ${\bm E}(t)$ and a static magnetic field ${\bm B}$, the equations of motion of the charge carriers, in a given band are given by~[\onlinecite{Chang1995, Sundaram1999, Gao2014, Gao2015}]
\bea \label{band_velocity}
\dot{\bm r} &=& D\left[\tilde {\bm v} + \dfrac{e}{\hbar}{\bm E} \times  {\bm \Omega} +\eta \dfrac{e}{\hbar}(\tilde{\bm v} \cdot {\bm \Omega}){\bm B}\right],
\\ \label{force}
\hbar \dot{\bm k} &=& D \left[-e {\bm E} - \alpha e(\tilde{\bm v} \times {\bm B}) - \zeta \dfrac{e^2}{\hbar}({\bm E}\cdot {\bm B}) {\bm  \Omega} \right].
\eea
Here, $ 1/D = \left[1 +\gamma \tfrac{e }{\hbar}{\bm B} \cdot {\bm \Omega} \right]$ is the phase-space  factor and ${\bm \Omega}={\bm \nabla}_{\bm k} \times \langle u | i {\bm \nabla}_{\bm k} | u \rangle$ is the Berry curvature with $|u \rangle$ being the periodic part of the Bloch wave-function. %Here and onward we omit the band and momentum index of Berry curvature for brevity. % 
The band velocity, modified by the orbital magnetic moment (OMM) ${\bm m}$, is given by $ \tilde {\bm v} =  {\bm v} - {\bm v}_{\rm m}$ where we have defined $\hbar {\bm v} = \partial \epsilon/\partial{\bm k}$ and $\hbar {\bm v}_{\rm m} = \partial \epsilon_{\rm m}/\partial{\bm k}$ with $\epsilon_{\rm m} = {\bm m} \cdot {\bm B}$. Note that the magnetic field modifies the band energy as $ \tilde{\epsilon} = \epsilon - \xi \epsilon_{\rm m} $, through the Zeeman like coupling of the magnetic field and the OMM, ${\bm m} = -i\frac{e}{2\hbar} \langle {\bm \nabla}_{\bm k} u | \times (\hat H - \epsilon)  |  {\bm \nabla}_{\bm k} u \rangle$. %
We emphasize that to keep track of the sources of various magnetic field dependences, we explicitly put $\alpha$ for the magnetic part of the Lorentz force, $\zeta$ for the `Berry force', $\eta$ for the magnetic velocity, $\gamma$ for the phase-space factor and $\xi$ for the OMM. At the end of the calculation, all these `tracking' factors will be set to 1.

The NDF is calculated from the iterative solutions of the Boltzmann equation within the relaxation time approximation. It is given by~[\onlinecite{Ashcroft1976}]
\begin{equation}\label{boltzmann}
\dfrac{\partial g (t)}{\partial t} + \dot{\bm k} \cdot {\bm \nabla}_{\bm k} g(t) =- \dfrac{g(t) - \tilde f}{\tau}.
\end{equation}
Here, $\tilde{f} = f(\epsilon - \xi {\bm m}\cdot{\bm B})$ with $f(\epsilon)=1/[e^{(\epsilon - \mu)/k_BT}]$ being the equilibrium Fermi-Dirac distribution function at chemical potential $\mu$ and temperature $T$ with $k_B $ being the Boltzmann constant. In Eq.~\eqref{boltzmann}, $\tau$ is the relaxation time whose energy dependence is ignored for simplicity.

In the weak electric field limit, the non-equilibrium part of the distribution function can be written as a power series of electric field dependent terms: $\sum_{n=1}^\infty f_n$ where $f_n \propto E^n$. We restrict ourselves upto $f_2$ for the calculation of second order NL response. Accordingly,  we consider the ansatz ~[\onlinecite{Deyo2009, Morimoto2016a}]
\be \label{ansatz}
f_2(t) = f_2^{(0)} + f_2^{(0)*}  + f_2^{(2\omega)} e^{i2\omega t} +  f_2^{(2\omega)*} e^{-i2\omega t}~,
\ee
where $f_2^{(0)}$ represents the DC (or rectification) part and $f_2^{(2\omega)} $ represents the second harmonic part of the NL distribution function. Substituting the ansatz of Eq.~\eqref{ansatz} into Eq.~\eqref{boltzmann} and following the usual Zener-Jones method~[\onlinecite{Hurd1972}], we obtain
\begin{equation}\label{dlta_n_2_m}
f_2^{(2\omega)} =\sum_{\nu=0}^{\infty}\left(\alpha D \tau_{2 \omega}\hat L \right)^\nu D\dfrac{e\tau_{2\omega}}{\hbar} \left({\bm E} + \zeta \dfrac{e}{\hbar} ({\bm E} \cdot {\bm B}){\bm  \Omega}\right)\cdot {\bm \nabla}_{\bm k} f_1^{(\omega)}.
\end{equation}
Here, $f_1^{(\omega)}$ is the first order in electric field contribution of the NDF [see Appendix \ref{linear} for details]. We have defined $
\hat L =  \tfrac{e }{\hbar}({\bm v} \times {\bm B})\cdot {\bm \nabla}_{\bm k}$ as the Lorentz force (magnetic part) operator. The modified scattering times are defined as $\tau_\omega = \tau/(1 + i\omega \tau) $ and $\tau_{2\omega} = \tau/(1 + i2\omega \tau) $. It is straightforward to expand the master solution, Eq.~\eqref{dlta_n_2_m}, and obtain the distribution function up to any  order of magnetic field dependence. In this paper we are interested in the lowest (linear) order of magnetic field dependence and the key steps of the calculation have been highlighted in Appendix~\ref{non-linear}. Note that from Eq.~\eqref{dlta_n_2_m} we can construct the rectification part of the distribution function $f_2^{(0)}$ by substituting $\tau_{2\omega} \to \tau$ and ${\bm E} \to {\bm E}^*$. 

%Now given the distribution function and the definition of current it is straight forward to calculate the NL conductivities. The general expressions of all the NL conductivities are given in the Appendix~\ref{non-linear}. 
%Here, we will enlist the conductivities irrespective of the underlying symmetry of the system. 

%{\bf expression of NL conductivities}

Using the calculated NDF, we now calculate the NL conductivities. We emphasize that in addition to the second order NL distribution function, the NL current can also arise from linear-${\bm E}$ part of the distribution function when it is combined with the anomalous velocity ${\bm E} \times {\bm \Omega}$. The NL conductivities can be expressed in the form of a momentum dependent conductivity, $\tilde \sigma_{ abc}$,  where $\sigma_{abc}= - e^3 \tau_{ \omega}/\hbar \int[d{\bm k}] \tilde \sigma_{ abc}$. The magnetic field independent part of the NL conductivity is obtained to be
\be \label{anmls_drd}
\tilde \sigma_{abc}^{(0)} =   \varepsilon_{abd} \Omega_d v_c f^\prime + \tau_{2\omega} v_a \partial_b v_c f^\prime ~.
\ee
Here, $\varepsilon_{abd}$ is the anti-symmetric Levi-civita symbol and $f'\equiv \partial_{\rm \epsilon} f$. The first term of Eq.~\eqref{anmls_drd} is the NL anomalous Hall conductivity~[\onlinecite{Sodemann2015}] and the second term is the NL Drude conductivity. 
The expression of magnetic field dependent part of the NL conductivity is a bit more complicated as various magnetic field contributions come into play. The total NL conductivity (linear in the magnetic field) can be expressed as $\tilde \sigma_{ abc}^{(\rm 1)}= \tilde \sigma_{ abc}^{( \eta)} + \tilde \sigma_{ abc}^{( \zeta)}  + \tilde \sigma_{ abc}^{( \alpha)} + \tilde \sigma_{ abc}^{( \gamma)}  + \tilde \sigma_{ abc}^{( \xi)}$. The different contributions to the magnetoconductivity can be calculated to be
\bea  \label{cmvlcty}
\tilde \sigma_{abc}^{(\eta)} &=&  \tau_{2\omega}  \Omega_{\rm v} B_a \partial_b v_c f^\prime,
\\[1ex] \label{bf}
\tilde \sigma_{abc}^{(\zeta)}  &=&  \varepsilon_{abd} \Omega_d \Omega_{\rm v} B_c f^\prime + \tau_{2\omega}  v_a(\Omega_{\rm B}^{db} \partial_d v_c + B_c \partial_b \Omega_{\rm v} ) f^\prime,
~~\\ [1ex]\label{lrntz}
\tilde \sigma_{abc}^{(\alpha)}  &=&  \tau_\omega \varepsilon_{abd}  \Omega_d \hat{L} v_c f^\prime + v_a \big( \tau_{2\omega}^2 \hat{L} \partial_b + \tau_\omega \tau_{2\omega} \partial_b \hat{L} \big) v_c f^\prime,
\\ [1ex] \label{phase-space}
\tilde \sigma_{abc}^{(\gamma)}  &=&  - \varepsilon_{abd} \Omega_d \Omega_{\rm B} v_c f^\prime - \tau_{2\omega}  v_a \left(\Omega_{\rm B} \partial_b + \partial_b \Omega_{\rm B} \right) v_c f^\prime,
\\[1ex] \label{omm}
\tilde \sigma_{abc}^{(\xi)}  &=&  -\varepsilon_{abd} \Omega_d \left( v_{{\rm m}c} f^\prime + \epsilon_{\rm m} v_c f^{\prime \prime} \right) - \tau_{2\omega}  v_{{\rm m}a} \partial_b v_c f^\prime \nn
\\ [1ex]
&& -  \tau_{2\omega}  v_a \partial_b (v_{{\rm m}c} f^\prime + \epsilon_{\rm m} v_c f^{\prime\prime}).
\eea
Here, we have defined $\Omega_{\rm B}^{db} \equiv (e/\hbar) \Omega_d B_b$ and $\Omega_{\rm v} \equiv (e/\hbar) {\bm \Omega} \cdot {\bm v}$ and $f''\equiv \partial_{\rm \epsilon}^2 f$. We emphasize here that the derivative operator $\partial_b=\partial/\partial k_b$ and $\hat L$ operate on all the terms appearing to their right hand side.
Together, Eqs.~\eqref{cmvlcty}-\eqref{omm} describe all the NL conductivity components, which vary linearly with the magnetic field. 

In materials which preserve SIS, the energy dispersion, Berry curvature and the OMM are even functions of the crystal momentum. %{\it i.e.} $\epsilon(-{\bm k})=\epsilon ({\bm k})$, ${\bm \Omega}(-{\bm k})={\bm \Omega}({\bm k})$ and ${\bm m}(-{\bm k})={\bm m}({\bm k})$. 
Consequently, the orbital magnetic moment coupling energy and the corresponding velocity satisfy $\epsilon_{\rm m}(-{\bm k})=\epsilon_{\rm m}({\bm k})$ and ${\bm v}_{\rm m}(-{\bm k})=-{\bm v}_{\rm m}({\bm k})$ respectively, and the bare band velocity obeys ${\bm v}(-{\bm k})=-{\bm v}({\bm k})$. Using these conditions, it is straightforward to show that all the NL conductivities [Eqs.~\eqref{anmls_drd}-\eqref{omm}] vanish in presence of SIS, as expected. 
%%Therefore, for finite second order NL responses the SIS of a material must be broken. 
If the SIS is broken, the presence or absence of TRS affects the NL conductivity. In presence of TRS, while the energy dispersion is an even function, the Berry curvature and the OMM are odd functions of the crystal momentum. Consequently we have $\epsilon_{\rm m}(-{\bm k})=-\epsilon_{\rm m}({\bm k})$, ${\bm v}_{\rm m}(-{\bm k})={\bm v}_{\rm m}({\bm k})$ and ${\bm v}(-{\bm k})=-{\bm v}({\bm k})$. Within these constraints, we find that for the magnetic field independent NL conductivities ($\sigma_{ abc}^{(\rm 0)}$), contributions that are quadratic in the scattering time vanish while linear scattering time dependent contributions are finite. On the other hand, for linear magnetic field dependent NL conductivity  ($\sigma_{ abc}^{(\rm 1)}$), contributions that are quadratic in scattering time survive while the other (linear and cubic) scattering time dependent contributions vanish.

For completeness, and to complement the discussion of the NL conductivity, we also discuss the linear conductivity of WSM [see Appendix \ref{linear} for detailed derivation]. The magnetic field independent linear conductivity is given by 
\be \label{sgm_Drd_anmls}
\sigma_{ab}^{(0)} = - e^2 \tau_\omega \int \left[ d{\bm k}\right] v_a v_b f^\prime  - \dfrac{e^2}{\hbar} ~ \varepsilon_{abc} \int \left[ d{\bm k}\right] \Omega_c f ~.
\ee
Here, the first term is the Drude conductivity and the second term is the intrinsic anomalous Hall conductivity which vanishes in TRS invariant systems. 
The linear order in magnetic field contribution to the linear conductivity is given by, 
\bea \label{sgm_bc_lrntz}
\sigma_{ab}^{(1)}  &=& -e^2 \tau_\omega \int [d{\bm k}] \Big[ \big(\eta \Omega_{\rm v} B_a v_b + \alpha \tau_\omega v_a \hat{L} v_b \nn
\\ 
&&  -\gamma v_a \Omega_{\rm B} v_b   + \zeta v_a \Omega_{\rm v} B_b \big) f^\prime - \xi \Big\{ \big( v_{{\rm m}a} v_b 
\\
&& + \hbar^{-1}\tau_\omega^{-1} \varepsilon_{abd}  \Omega_d \epsilon_{\rm m}  +  v_a v_{{\rm m}b} \big ) f^\prime 
 + v_a \epsilon_{\rm m} v_b f^{\prime\prime} \Big\} \Big].~~ \nn
\eea
It can be easily checked that in a TRS invariant system, the diagonal components ($a=b$) of Eq.~\eqref{sgm_bc_lrntz} will vanish, consistent with Onsager's relation. So we can only have the linear-${\bm B}$ dependent Hall components ($a\neq b$). Furthermore, we find that in presence TRS only those contributions to the linear-${\bm B}$  conductivity are nonzero which are quadratic and zeroth order in the  scattering time. From Eq.~\eqref{sgm_bc_lrntz} it is evident that the quadratic dependence of scattering time is described by the Lorentz force and the scattering time independent contribution has its origin in the OMM~[\onlinecite{Das2021}].

The framework presented in this section for exploring the NL magneto-conductivity is very general, and applicable to all SIS broken materials. In the rest of the paper, we apply this to explore the NL magneto-conductivity in time reversal symmetric WSM. 
\begin{figure}[t!]
\includegraphics[width =1.0\linewidth]{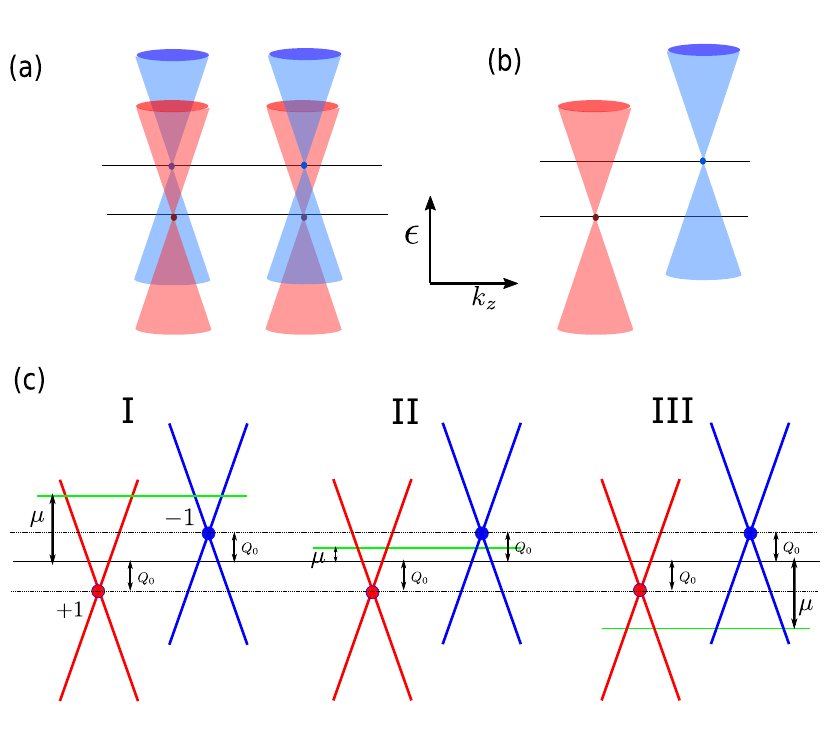}
\caption{A schematic depiction of the band dispersion for a) TRS preserving and SIS broken WSM, and b) TRS broken (with SIS preserved) WSM. Due to the presence of TRS, minimum four Weyl nodes are required in the SIS broken WSM. In contrast, when TRS is absent and SIS is present, a minimum two Weyl nodes is possible. (c) shows three different scenarios for the location of the Fermi level, with respect to the location of the Weyl points. In the left panel the Fermi level is in the conduction band of both Weyl nodes ($\mu > Q_0 > 0$). In the middle panel the Fermi energy lies in the conduction band of one Weyl node and in the valence band of the other Weyl node ($Q_0 > \mu > -Q_0$). The right panel shows the scenario when the Fermi level lies in the valence band of both the Weyl nodes ($\mu < -Q_0 < 0$).
\label{fig_2}}
\end{figure}

\section{Nonlinear conductivity in Weyl semimetal}
\label{model_WM}
%

%{\bf implications of symmetries in WSM in general}
In this section, we calculate the linear and NL magneto-conductivity for an SIS broken WSM. %For the calculation we consider two different cases of WSM: a) WSM with the broken SIS but preserved TRS, and b) WSM where both the TRS and SIS are broken. 
%We emphasize here that since we are interested in the second order NL response, the SIS of the considered system must be broken. 
The simultaneous presence of TRS and SIS forces all the bands in the given material to be doubly generate, and this excludes the possibility of the formation of a WSM, in which two non-degenerate linearly dispersing bands cross each other [\onlinecite{Amritage2018}]. Thus, for realizing a WSM state, either SIS or TRS or both the symmetries must be broken. 
%Any one among these (or both) symmetries must be broken~[\onlinecite{Amritage2018}]. 
%Now we discuss implications of these symmetries and constraints imposed on the low energy model. 
In a TRS preserving (SIS broken) WSM, a minimum of four Weyl nodes~[\onlinecite{Nielsen1983, Murakami2007, Belopolski2017}] have to be there. Among the four nodes, the nodes with the same chirality are connected by a time reversal invariant momentum and corresponding charge neutrality points reside at the same energy. However, there are no symmetry restriction among the nodes with different chirality. On the other hand, in a SIS preserving (TRS broken) WSM, a minimum of a single pair of two Weyl nodes of opposite chirality are allowed, and for each such pair of Weyl nodes, the corresponding charge neutrality point reside at different energies. Both of these scenarios have been sketched in Fig.~\ref{fig_2}(a)-(b) respectively. In our work, we present the NL conductivity calculation for a TRS invariant WSM with minimum four Weyl nodes shown in Fig.~\ref{fig_2}(a).

The low energy Hamiltonian of a single Weyl node can be written as~[\onlinecite{Zyuzin2012,Vazifeh2013, Chang2015, Tabert2016, Mukherjee2018}]
\be \label{H_Weyl}
H_s ( {\bm k}) = s\hbar v_F {\bm \sigma} \cdot ({\bm k} - s {\bm Q}) - sQ_0~.
\ee
Here, ${\bm k}$ is the crystal momentum, ${\bm \sigma}=(\sigma_x, \sigma_y, \sigma_z)$ are the Pauli matrices, $v_{F}$ is the Fermi velocity and $s = \pm 1$ is the chirality index.
In Eq.~\eqref{H_Weyl}, the ${\bm Q}$ and $Q_{0}$ denotes the position of the Weyl nodes in the momentum and energy, respectively. For simplicity of calculation, we consider the case where the Weyl node is situated at the origin and emphasize that non-zero ${\bm Q}$, which breaks the time reversal symmetry, does not alter any of the results. With ${\bm Q} =0$, the  Hamiltonian in Eq.~\eqref{H_Weyl} simplifies to, $H_s ({ \bm k}) = s \hbar v_F {\bm \sigma} \cdot {\bm k} - s Q_0$.
The breaking of SIS in this model is reflected through the finite $Q_0$ which positions the Weyl nodes of opposite chirality at different energy. More specifically, the Weyl point of the positive chirality node lie at energy $-Q_{\rm 0}$ while the Weyl point for the negative chirality node lies at $Q_{\rm 0}$, making the energy separation between the two Weyl nodes to be $2Q_0$. On the other hand, the TRS is enforced here by considering a  minimum of four Weyl nodes in such a way that the two nodes of same chirality out of the four are situated at the same energy [see Fig.~\ref{fig_2}(a)]. The energy dispersion and band velocity for the Hamiltonian are given by $\epsilon_s({\bm k}) = s_{\rm b} \hbar v_F k - s Q_{\rm 0}$ and  ${\bm v}({\bm k}) = s_{\rm b} v_F {\bm k}/k$ respectively, where $k = |{\bm k}|$ and $s_{\rm b} = +1 (-1)$ for the conduction (valence) band. We note that non-zero $Q_0$ makes $\epsilon_+ ({\bm k}) \neq \epsilon_-({\bm k})$, implying the breaking of inversion symmetry. The Berry curvature and the OMM for this model are given by~[\onlinecite{Morimoto2016c, Li2021}]
\be
{\bm \Omega} = -s s_{\rm b} \dfrac{\bm k}{2k^3} ;~~~
{\bm m}= -s e v_F \dfrac{\bm k}{2k^2},
\ee
respectively. We note that these band geometric quantities do not depend on the energy separation ($Q_0$) of the Weyl points in the corresponding Weyl nodes. In contrast to  the Berry curvature, the OMM is independent of the band index. Both these band geometric quantities are highly concentrated near the Weyl point, as expected.

\setlength{\tabcolsep}{7.5pt}
\renewcommand{\arraystretch}{1.7}
\begin{table*}
%\captionsetup{format=plain,labelsep=period,width=\textwidth}
\caption{Symmetrized nonlinear conductivity  $\bar{\sigma}_{abc}$ in the unit of $\tilde{\sigma}_{\rm NL}$, which is defined in Eq.~\eqref{unit}, for the TRS invariant WSM. The Greek indices in the parentheses denote the sources of the magnetic field that are inducing finite conductivity and the total is obtained by adding all the contributions. The diagonal components of the NL conductivity are identically zero. The NL Hall conductivities with same last two indices ($\sigma_{abb}$) are non-zero when $a$ and $b$ are cyclic coordinates and the magnetic field dependence is dictated by $B_a$. The Hall conductivities with different last two indices can be written as $\bar \sigma_{aab}=\bar \sigma_{aba}$ and its magnetic field dependence is determined by $B_b$. \label{table_NLC_isbroken_v3}}
\begin{tabular}{ccccccc}
\hline \hline
$\bar{\sigma}_{abc}$ & $\bar{\sigma}_{axx}$ & $\bar{\sigma}_{ayy}$ & $\bar{\sigma}_{azz}$ & $\bar{\sigma}_{axy}$ & $\bar{\sigma}_{ayz}$ & $\bar{\sigma}_{azx}$
\\ 
\hline\hline
$\bar{\sigma}_{xbc}$ & 0 & $ B_x (2\eta + \alpha - \gamma) $ & 0 & $ -B_y (\frac{\alpha}{2} - \frac{\gamma}{2} + \zeta) $ & 0 & 0
\\
\hline
$\bar{\sigma}_{ybc}$ & 0 & 0 & $ B_y (2\eta + \alpha - \gamma) $ & 0 & $ -B_z (\frac{\alpha}{2} - \frac{\gamma}{2} + \zeta) $ & 0 
\\
\hline
$\bar{\sigma}_{zbc}$ & $ B_z (2\eta + \alpha - \gamma) $ & 0 & 0 & 0 & 0 & $ -B_x (\frac{\alpha}{2} - \frac{\gamma}{2} + \zeta) $
\\
\hline\hline
\end{tabular}
\end{table*}
%

%{\bf results summarized from the table}

Using these in Eqs.~\eqref{anmls_drd}-\eqref{omm}, 
we calculate the NL conductivity and the symmetrized results $\bar{\sigma}_{abc}=(\sigma_{abc}+\sigma_{acb})/2$  are summarized in Table ~\ref{table_NLC_isbroken_v3}. We note that the NL Drude conductivity is identically zero due to the presence of TRS. Furthermore, we find that the NL anomalous Hall conductivity is also zero. Although individual Weyl node possesses finite NL anomalous Hall conductivity, the total response vanishes after summing over the nodes. This happens due to the absence of Berry curvature dipole in this system. In order to realize the  NL anomalous response~[\onlinecite{Sodemann2015}], the mirror symmetry has to be broken. This is generally  achieved in WSM with tilt or with higher order (band  bending) terms~[\onlinecite{Sodemann2015,Facio2018,Matsyshyn2019,Ortix2021}] in the effective Hamiltonian. 

Coming to the linear magnetic field dependent NL conductivities, all the conductivities and their origin of magnetic field dependences are explicitly highlighted in Table~\ref{table_NLC_isbroken_v3}. For compactness, the various components of the conductivity are expressed as  $\bar{\sigma}_{abc}=\tilde \sigma_{\rm NL}B_d (\eta, \alpha, \gamma, \zeta, \xi)$ where $B_d$ is the component of the magnetic field along the $d \in (x,y,z)$ axis, and the total contribution is given by the sum of different magnetic field sources. We have defined
\be \label{unit}
\tilde \sigma_{\rm NL}= \dfrac{e^4 v_F^2 \tau^2}{\pi^2 \hbar^2 |\mu|} \frac{r_0}{3(1 - r_0^2)}~,
\ee
with $r_0 \equiv Q_0/|\mu|$. We find three key features in the NL conductivities. i) All the longitudinal NL conductivities ($\sigma_{aaa}$) vanish within the linear-${\bm B}$ approximation. ii) We find that the NL Hall components with the same last two indices, which we term as `pure' Hall components, such as $\bar{\sigma}_{zxx}, \bar{\sigma}_{yzz}, \bar{\sigma}_{xyy}$ are non-zero and determined by the magnetic velocity term  ($\eta$) in addition to the phase-space factor ($\gamma$) and the magnetic part of the Lorentz force ($\alpha$). These NL conductivities can be expressed as $\sigma_{abb} \propto B_a \delta_{{\hat {\bm a}} \times {\hat {\bm b}}, {\hat {\bm c}}}$ which implies that the currents corresponding to the conductivity flow along the direction of the applied magnetic field. iii) The NL Hall components with different the last two indices, which we term as the `mixed' Hall components, are determined by the Berry force ($\zeta$) in addition to the phase-space factor ($\gamma$) and the Lorentz force ($\alpha$). The magnetic field dependence of the mixed components, $\sigma_{aab} \propto B_b$, implies that the currents corresponding to the conductivity flow perpendicular to the magnetic field direction. These are the main findings of our paper in the context of SIS broken WSM. Interestingly, we find that the OMM ($\xi$) contributions to the NL conductivity is  identically zero for the TRS preserving case.

We emphasize here that for the calculation of conductivities for Weyl nodes with opposite chirality separated in energy, three different scenarios based on the position of the Fermi level as highlighted in Fig.~\ref{fig_2}(c), are possible. In scenario-I, the Fermi level resides in the conduction band of both the Weyl nodes. In  scenario-II, the Fermi level resides in the conduction  band of one Weyl node and in the valence band of the other  Weyl node. In scenario-III, the Fermi level resides in the valence band of both the Weyl nodes. Interestingly, we find that the total conductivity, including all the Weyl nodes, can be expressed by same expression for these three scenarios. This has been shown in Appendix~\ref{istrpc_WM} in detail.

%Note that the occupancy at each Weyl node is different  depending on the position of the Fermi level [see Fig.~\ref{fig_2}(c)], due to the energy difference between the Weyl points of the opposite chirality nodes. For that we have considered three different scenarios based on the position of the Fermi level, as highlighted in Fig.~\ref{fig_2}(c). In scenario-I, the Fermi level resides in the conduction band of both the Weyl nodes. In  scenario-II, the Fermi level resides in the conduction  band of one Weyl node and in the valence band of the other  Weyl node. In scenario-III, the Fermi level resides in   the valence band of both the Weyl nodes. We find that the total conductivity, including all the Weyl nodes, can be expressed by same expression for these three scenarios as  shown in Appendix~\ref{istrpc_WM} in detail.

Although we have considered the three scenarios, a continuous transition of Fermi level among them is not allowed within our formalism. This is due to the fact that the employed semiclassical Boltzmann formalism is generally valid at high carrier densities such that $\mu \gg \hbar/\tau$ ~[\onlinecite{Ashcroft1976}]. Hence, the above results are not applicable near the limit $|r_{\rm 0}| = 1$, where the chemical potential is located at the Weyl point of either of the two Weyl nodes. 

%s{\bf comparison with literature}

Now, we compare our results with some recent related works~[\onlinecite{Morimoto2016c,Li2021}] and highlight the differences. We note that in  Ref.~[\onlinecite{Morimoto2016c}] the authors provide result of a single Weyl node. To our satisfaction, we can obtain those results by putting $Q_0=0$ in our single node calculations presented in Appendix~\ref{istrpc_WM}. We emphasize that total NL responses from all nodes are zero for $Q_0=0$ as the inversion symmetry is restored. 
Recently in Ref.~[\onlinecite{Li2021}], it has been shown that the inversion symmetry broken tilted WSM possesses NL magnetoresponse. We note that the NL conductivities discussed in Ref.~[\onlinecite{Li2021}] are linear in scattering time, $\tau$, however the NL conductivities we discuss in our paper are proportional to the square of the scattering time, $\tau^2$. We emphasize here that the TRS has to be broken in addition to SIS in order to obtain linear-$\tau$ dependent NL magneto-conductivity. Comparing the order of magnitude of NL conductivities discussed in our paper with the results in Ref.~[\onlinecite{Li2021}] we find that for certain parameter values $v_{\rm F}=3\times {\rm 10^5}$ m/s, $\tau={\rm 10^{-13}}$ s, $R_{\rm s}(\text{tilt})={\rm 0.5}$, $\mu={\rm 20}$ meV and $Q_{\rm 0}={\rm 10}$ meV, the conductivities are of the comparable order: $\sigma(R_{\rm s}={\rm 0})/\sigma(R_{\rm s} \neq {\rm 0}) \sim \mathcal{O}({\rm 10^1})$, where $\sigma(R_{\rm s}={\rm 0})=\tilde{\sigma}_{\rm NL}$ and $\sigma(R_{\rm s} \neq {\rm 0})$ is the NL conductivity in Eq.~({\rm 11}) of [\onlinecite{Li2021}].

%{\bf linear conductivities}

For completeness, we now discuss the linear conductivity  of WSM, starting from Eqs.~\eqref{sgm_Drd_anmls}-\eqref{sgm_bc_lrntz}. As expected from the Onsager relations, the magnetic field induced part of the longitudinal components are identically zero and the Drude conductivity is obtained to be~[\onlinecite{Tabert2016}]
\be  \label{sgm_aa_isbrk_rslt}
\sigma_{aa} = \dfrac{2 e^2 \tau \mu^2 (1 + r_0^2)}{3 \pi^2 \hbar^3 v_F}~,
\ee
where $a=(x,y,z)$. The Drude contribution is linear in the  scattering time and quadratic in the node separation. The  Drude conductivity in Eq~\eqref{sgm_aa_isbrk_rslt} reduces to the known result~[\onlinecite{Tabert2016, Das2019a}] in the limit of zero energy separation between the Weyl points, $Q_{\rm 0}={\rm 0}$. Here, the modification in the Drude conductivity corresponds to the fact that the energy separation between the nodes gives rise to different carrier concentrations at the opposite chirality. 
The linear Hall conductivity is given by
\be \label{sgm_ab_isbrk_rslt}
\sigma_{ab} = - \varepsilon_{abc} B_c\dfrac{ e^3 v_F \mu}{6 \pi^2 \hbar^3} \left(\xi \dfrac{\hbar^2}{\mu^2 (1 - r_0^2)} + 4\alpha \tau^2 \right)~.
\ee
%
%where the indices $a,b,c$ are in cyclic order with $\sigma_{ba}=-\sigma_{ab}$. \textcolor{red}{Here you probably need to use $\epsilon_{abc}$ - the Levi cavita tensor, else it does not make sense. Why can't a=b}. 
We highlight that the Hall conductivity has two components, one is quadratic and the other is zeroth order in the scattering time. While the former extrinsic contribution  has its origin in the Lorentz force (term $\propto \alpha$), the latter intrinsic contribution (term $\propto \xi$) originates from the OMM~[\onlinecite{Das2021}].

%\section{Discussion} \label{discussion}
%
%A finite NL Hall current is found in inversion broken WM in absence of any tilt in the dispersion. We have discussed the emergence of NL charge response in two different kind of Weyl phases considering all the possible sources of magnetic field appearing in the equation of motion. It has been shown that the Hall components are linearly dependent on ${\rm B}$, while it is absent in the longitudinal conductivities. Further the role of Weyl node separation in generating nonlinear Hall current, in both momentum and energy space, has also been illustrated. 
%%

%%%%%%%%%%%%%%%%%%%%%%%%%%%%%%%%%%%%%%%%%%%%%%%%%%%%%%%

\section{Summary and Conclusion}\label{conclusion}
To wrap up, in this paper we have investigated the second order NL conductivity in time reversal symmetric WSM in presence of a weak magnetic field. Starting from the semiclassical Boltzmann transport formalism, we obtain the general expressions of all the NL Hall conductivities, and identify the different physical mechanisms contributing to the magnetic field dependencies. Using the developed framework for NL magneto-conductivity in conjugation with appropriate symmetry  analysis, we calculate the NL Hall conductivity in WSM without any tilt. Our calculations explicitly highlight the interplay of the quantum geometric Berry curvature and the  magnetic part of the Lorentz force. 
We predict two types of new NL Hall effects. In one case, which we term as pure NL Hall effect, the current flows along the direction of the magnetic field, but perpendicular to the applied electric field. In the other case, which we term as the mixed NL Hall effect, the NL current flows perpendicular to the magnetic field direction. These newly predicted NL conductivity in WSM can be probed  through NL magneto-transport or through NL magneto-optical experiments.

\section{ACKNOWLEDGMENTS}
We acknowledge Department of Physics, IIT Kanpur, the Science Education and
Research Board (SERB) and the Department of Science and
Technology (DST), Government of India.

\appendix
\section{Semiclassical theory for linear conductivities}
\label{linear}
In this section of Appendix, we will calculate current that is linear in ${\bm E}$-field and linear in ${\bm B}$-field. We assume that in the steady state the first order distribution function oscillates with fundamental frequency with the form~[\onlinecite{Deyo2009, Morimoto2016c}]
$f_1( t) = f_1^{(\omega)} e^{i\omega  t} + f_1^{(\omega)*} e^{-i\omega t}$. Using this ansatz in the Boltzmann equation, Eq.~\eqref{boltzmann}, we obtain
\be \label{dlta_n_1}
f_1^{(\omega)} =\sum_{\nu} \left(\alpha D \tau_\omega  \hat L\right)^\nu D\dfrac{e \tau_\omega }{\hbar} \left[ {\bm E} + \zeta \dfrac{e}{\hbar} \left( {\bm E} \cdot {\bm B} \right) {\bm \Omega}\right] \cdot {\bm \nabla}_{\bm k} \tilde f ~.
\ee
We expand the series in Eq.~\eqref{dlta_n_1} in orders of ${\bm B}$-field in the limit of small magnetic field~[\onlinecite{Hurd1972, Gao2019, Cortijo2016}]. Using this expansion, the non-equilibrium part can be written as $f_1^{(\omega)} = f_{10}^{(\omega)} + f_{11}^{(\omega)}$, where the first subscript denotes the order of electric field and the second subscript denotes the order of magnetic field. We obtain
\small
\bea \label{dlta_n_1_0}
f^{(\omega)}_{10} &=& e \tau_\omega {\bm E} \cdot  {\bm v} f^\prime ~,
\\  \nn
f^{(\omega)}_{11}   &=&  e\tau_\omega (\zeta \Omega_{\rm v} {\bm B}
- \gamma \Omega_{\rm B}  {\bm v}
)\cdot  {\bm E} f^\prime 
- \xi e\tau_\omega  {\bm E} \cdot \left({\bm v}_{\rm m} f^\prime + \epsilon_{\rm m} {\bm v} f^{\prime \prime} \right) 
\\ \label{dlta_n_1_1}
&&+ \alpha e \tau_\omega^2  \hat L   {\bm v}\cdot  {\bm E} f^\prime ~.
\eea
We note that these results are consistent with the previous studies~[\onlinecite{Morimoto2016c, Cortijo2016,Das2019b, Gao2019}].

Using the distribution functions we can calculate current. The zeroth order in ${\bm B}$-field current can be written as ${\bm j}_{10}( t) = {\bm j}_{10}^{(\omega)} e^{i \omega t} + {\bm j}_{10}^{(\omega)*} e^{-i \omega  t} $. Separating the different order of scattering time dependence, we obtain
\bea \label{lin_current_B0}
{\bm j}_{10}^{(\omega)} (\tau_\omega^0) &=& -\dfrac{e^2}{\hbar} \int [d{\bm k}] ({\bm E} \times {\bm \Omega}) f~,
\\
{\bm j}_{10}^{(\omega)} (\tau_\omega) &=& - e^2 \tau_\omega \int[d{\bm k}] {\bm v}({\bm E} \cdot {\bm v}) f^\prime~.
\eea
The first one is the intrinsic anomalous Hall effect where current flows perpendicular to the electric field. Symmetry analysis shows that in presence of TRS, the anomalous Hall effect  vanishes. The second one is the ordinary Drude current. 
%Its component along the electric field (${\bm E} \cdot {\bm j}$) is always non-zero. However for rotational invariant systems, the perpendicular component (${\bm E} \times {\bm j}$) is zero due to Onsager's reciprocal relation $\sigma_{ab} = \sigma_{ba} ~(i \neq j)$. 
The linear order in ${\bm B}$-field current can be written as ${\bm j}_{11}( t) = {\bm  j}_{11}^{(\omega)} e^{i \omega  t} + {\bm  j}_{11}^{(\omega)*} e^{-i \omega  t}$. We emphasize that `equilibrium' distribution function in presence of a magnetic field, $\tilde f  = f - \xi \epsilon_{\rm m}  f^\prime$ also contributes to the current in addition to the non-equilibrium parts Eqs.~\eqref{dlta_n_1_0}-\eqref{dlta_n_1_1}. Now, using these we calculate current in various order in scattering time. The even power of scattering time dependent current is given by
\bea \label{lin_current_B1}
{\bm j}_{11}^{(\omega)} (\tau_\omega^0)&=& \dfrac{e^2}{\hbar}\xi  \int [d{\bm k}] \left({\bm E} \times {\bm \Omega}\right) \epsilon_{\rm m} f^\prime ,
\\ \label{lin_current_B2}
{\bm j}_{11}^{(\omega)} (\tau_\omega^2) &=&  -e^2 \tau_\omega^2\alpha  \int [d{\bm k}] {\bm v} \hat L   {\bm v}
\cdot  {\bm E} f^\prime~.
\eea 
Current that is linear order in scattering time is given by
\begin{equation} \label{lin_current_B3}
\begin{aligned}
{\bm j}_{11}^{(\omega)} (\tau_\omega) =  e^2 \tau_\omega \int  [d{\bm k}] \Big[ \xi {\bm v}_{\rm m} {\bm E} \cdot {\bm v} f^\prime -  \eta  {\bm  B}  \Omega_{\rm v}  {\bm v} \cdot  {\bm E} f^\prime - 
\\
{\bm v} \left(\zeta \Omega_{\rm v} {\bm B}
- \gamma \Omega_{\rm B}  {\bm v} 
\right)\cdot  {\bm E} f^\prime~
+ 
 \xi {\bm v} {\bm E} \cdot \left({\bm v}_{\rm m} f^\prime + \epsilon_{\rm m} {\bm v} f^{\prime \prime} \right)\Big].
\end{aligned}
\end{equation}
The linear-${\bm B}$ dependent currents calculated in Eqs.~\eqref{lin_current_B1}-\eqref{lin_current_B3} have been earlier discussed in Refs.~[\onlinecite{Cortijo2016, Morimoto2016c,Das2019b,Gao2019}].
In presence of TRS (broken SIS) various quantities satisfy $(\epsilon_{\rm m}, {\bm \Omega})(-{\bm k})= -(\epsilon_{\rm m}, {\bm \Omega})({\bm k})$, ${\bm v}(-{\bm k}) = -{\bm v}({\bm k})$ and ${\bm  v}_{\rm m}(-{\bm  k}) = {\bm v}_{\rm m}({\bm k})$, hence all the contributions $\propto \tau_\omega$ vanish. However, currents proportional to the even power of scattering time survives, out of which the Lorentz force contribution $\propto \tau_\omega^2$ gives rise to the classical Hall effect~[\onlinecite{Hurd1972, Ziman1972}], and the anomalous velocity contribution $\propto \tau_\omega^0$ gives rise to OMM induced intrinsic Hall effect~[\onlinecite{Das2021}]. On the other hand, in presence of SIS (broken TRS) the various quantities satisfy $(\epsilon_{\rm m}, {\bm \Omega})(-{\bm k})= (\epsilon_{\rm m}, {\bm \Omega})({\bm k})$, ${\bm v}(-{\bm k}) = -{\bm v}({\bm k})$,  ${\bm v}_{\rm m}(-{\bm k}) = -{\bm v}_{\rm m}({\bm k})$, and in that case all the linear-${\bm B}$ dependent terms are expected to be non-zero. 

\section{Semiclassical theory for the nonlinear conductivities}
\label{non-linear}
In this section of Appendix, we calculate current that is NL in ${\bm E}$-field. The ansatz for the non-equilibrium part of the distribution function quadratic in ${\bm E}$-field~[\onlinecite{Deyo2009, Morimoto2016c}] can be written as 
\be
f_2(t) = f_2^{(0)} + f_2^{(0)*}  + f_2^{(2\omega)} e^{i2\omega t} +  f_2^{(2\omega)*} e^{-i2\omega t}~,
\ee
where $f_2^{(0)}$ represents the rectification part and $f_2^{(2\omega)} $ represents the second harmonic part. With this, from Eq.~\eqref{boltzmann} we obtain
\begin{equation}\label{dlta_n_2}
f_2^{(2\omega)} =\sum_\nu \left(\alpha D \tau_{2 \omega}\hat L\right)^\nu D\dfrac{e\tau_{2\omega}}{\hbar} \left[{\bm E} + \zeta \dfrac{e}{\hbar} ({\bm E} \cdot {\bm B}){\bm \Omega}\right]\cdot {\bm \nabla}_{\bm k} f_1^{(\omega)},
\end{equation}
and 
\be\label{dlta_n_2_dc}
f_2^{(0)} =\sum_\nu \left(\alpha D \tau \hat L\right)^\nu D\dfrac{e\tau}{\hbar} \left[{\bm E}^*+ \zeta \dfrac{e}{\hbar} \left({\bm E}^* \cdot {\bm B}\right) {\bm \Omega}  \right]\cdot {\bm \nabla}_{\bm k} f_1^{(\omega)} .
\ee
We emphasize that from Eq.~\eqref{dlta_n_2} we can generate Eq.~\eqref{dlta_n_2_dc} by $\tau_{2\omega} \to \tau$ and ${\bm E} \to {\bm E}^*$. From this master solution it is now straightforward to separate out the distribution function in various order in magnetic field as $ f_2^{(2\omega)} = f_{20}^{(2\omega)} + f_{21}^{(2\omega)}$. These can be calculated as
\begin{widetext}
\begin{eqnarray}\label{dlta_n_2_0}
f^{(2\omega)}_{20} (\tau_{2\omega}, \tau_\omega)&=& \dfrac{e^2 \tau_{2\omega} \tau_\omega}{\hbar} ({\bm E}\cdot {\bm \nabla}_{\bm k} ){\bm E}\cdot {\bm v} f^\prime~,
\\\nn
f^{(2\omega)}_{21} (\tau_{2\omega}, \tau_\omega)&=& \dfrac{e^2 \tau_{2\omega}\tau_\omega}{\hbar} 
 \big [ \zeta \dfrac{e}{\hbar} ({\bm E} \cdot {\bm B})( {\bm \Omega} \cdot {\bm \nabla}_{\bm k} ){\bm v}\cdot {\bm E} f^\prime - \gamma \Omega_{\rm B}  ({\bm E}  \cdot {\bm \nabla}_{\bm k} ) {\bm v}\cdot {\bm E} f^\prime + ({\bm E} \cdot {\bm \nabla}_{\bm k}) \left\lbrace \left(\zeta \Omega_{\rm v} {\bm B} - \gamma \Omega_{\rm B}  {\bm v} \right)\cdot  {\bm E} f^\prime \right.
\\\label{dlta_n_2_1} 
&& \left. - \xi  {\bm E} \cdot \left({\bm v}_{\rm m} f^\prime + \epsilon_{\rm m} {\bm v} f^{\prime \prime} \right) \right\rbrace
\big]~,
\\ \label{f_tau_3_1}
f^{(2\omega)}_{21} (\tau_{2\omega}^2, \tau_\omega)&=& \alpha\dfrac{e^2 \tau_{2\omega}^2 \tau_\omega}{\hbar} \hat L  ({\bm E}  \cdot {\bm \nabla}_{\bm k}) {\bm E} \cdot {\bm v} f^\prime,~~~~~~~~~
f^{(2\omega)}_{21} (\tau_{2\omega}, \tau_\omega^2) = \alpha\dfrac{e^2 \tau_{2\omega} \tau_\omega^2}{\hbar}  ({\bm E}  \cdot {\bm \nabla}_{\bm k}) \hat L {\bm v} \cdot {\bm E}  f^\prime~.
\end{eqnarray}
\end{widetext}
We note that the first two expressions are $\propto \tau^2$ and the last two expressions are $\propto \tau^3$. We can obtain the rectification part, $f_2^{(0)}$ from these expressions just by replacing $\tau_{2\omega}$ by $\tau$ and ${\bm E}$, first one from the left, by ${\bm E}^*$. These results are consistent with the previous studies~[\onlinecite{Cortijo2016, Morimoto2016c}].

%Now, we will look for the current expression that is quadratic in ${\bm E}$-field and up to linear order in ${\bm B}$-field. Different magnetic field dependences in the charge current will be discussed separately.

The magnetic field independent contributions to the current come from the semiclassical band velocity and Berry curvature induced anomalous velocity. The second harmonic current that is zeroth order in magnetic field, can be written as ${\bm j}_{20}(t) = {\bm j}_{20}^{(2\omega)} e^{i 2\omega  t} + {\bm j}_{20}^{(2\omega)*} e^{-i 2\omega t}$ where 
\bea
{\bm j}^{(2\omega)}_{20} (\tau_\omega) &=& - \dfrac{e^3 \tau_\omega}{\hbar}  \int[d{\bm  k}]({\bm  E} \times {\bm \Omega}) {\bm v}\cdot{\bm E} ~f^\prime,
\\
{\bm j}^{(2 \omega)}_{20} (\tau_\omega, \tau_{2\omega}) &=& -\dfrac{ e^3 \tau_\omega\tau_{2\omega} }{\hbar}\int [d{\bm k}] {\bm v} ~({\bm E} \cdot {\bm  \nabla}_{\bm k}) {\bm E} \cdot {\bm v} f^\prime.~~~
\eea
The rectification part can be written as ${\bm
 j}_{20} = {\bm j}_{20} + {\bm j}_{20}^{*}$ replacing  $\tau_{2\omega}$ by $\tau$ and ${\bm E}$ (first one from the left) by ${\bm E}^*$ in the above expressions.
The ${\bm j}^{(2\omega)}_{20} (\tau_\omega)$ represents the NL anomalous current~[\onlinecite{Sodemann2015,Kang2019}] while the second term is the ordinary NL Drude current originating from the band velocity. In presence of TRS but broken SIS, only the NL anomalous Hall contribution is expected to be non-zero. The breaking of SIS plays the key role in the quadratic NL response, as in presence of SIS both the contributions vanish identically.

\squeezetable
\setlength{\tabcolsep}{2pt}
\renewcommand{\arraystretch}{2.5}
\begin{table*}[t]
%\captionsetup{format=plain,labelsep=period,width=\textwidth}
\begin{tabular}{cccccccccc}
\hline \hline
$\sigma_{abc}$ & $\sigma_{axx}$ & $\sigma_{axy}$ & $\sigma_{axz}$ & $\sigma_{ayx}$ & $\sigma_{ayy}$ & $\sigma_{ayz}$ & $\sigma_{azx}$ & $\sigma_{azy}$ & $\sigma_{azz}$
\\
\hline\hline
$\sigma_{xbc}$ & $ (-2\eta + 2\zeta) B_x $ & $ (-\gamma + 2\zeta) B_y $ & $ (-\gamma + 2\zeta) B_z $ & $ \alpha B_y $ & $ (-2\eta - \alpha + \gamma) B_x $ & $ -s_{\rm b} \dfrac{\sigma_0}{\chi_0} $ & $ -\alpha B_z $ & $ -s_{\rm b} \dfrac{\sigma_0}{\chi_0} $ & $ (-2\eta + \alpha + \gamma) B_x $ 
\\
\hline
$\sigma_{ybc}$ & $ (-2\eta + \alpha + \gamma) B_y $ & $ -\alpha B_x $ & $ -s_{\rm b} \dfrac{\sigma_0}{\chi_0} $ & $ (-\gamma + 2\zeta) B_x $ & $ (-2\eta + 2\zeta) B_y $ & $ (-\gamma + 2\zeta) B_z $ & $ -s_{\rm b} \dfrac{\sigma_0}{\chi_0} $ & $ \alpha B_z $ & $ (-2\eta - \alpha + \gamma) B_y $
\\
\hline
$\sigma_{zbc}$ & $ (-2\eta - \alpha + \gamma) B_z $ & $ -s_{\rm b} \dfrac{\sigma_0}{\chi_0} $ & $ \alpha B_x $ & $ -s_{\rm b} \dfrac{\sigma_0}{\chi_0} $ & $ (-2\eta + \alpha + \gamma) B_z $ & $ -\alpha B_y $ & $ (-\gamma + 2\zeta) B_x $ & $ (-\gamma + 2\zeta) B_y $ & $ (-2\eta + 2\zeta) B_z $
\\
\hline\hline
\end{tabular}
\caption{Nonlinear conductivity elements (in unit of $s s_{\rm b} \chi_0/12$) for a single node of a WSM with band crossing (charge neutrality point) situated at zero energy. Here, we have defined $\chi_0 = e^4 \tau^2 v_F^2/(\pi^2 \hbar^2 |\mu|)$ and $\sigma_0 = e^3\tau/(\pi^2\hbar^2) $. Note that we have also included the NL anomalous Hall effect here which are independent of magnetic field.}
\label{table_NLC_single}
\end{table*}

For the linear-${\bm B}$ contributions, the second harmonic current can be similarly written as ${\bm
 j}_{21}( t) = {\bm j}_{21}^{(2\omega)} e^{i 2\omega  t} + {\bm j}_{21}^{(2\omega)*} e^{-i 2\omega  t}$. In various orders of scattering time we obtain
\begin{widetext}
\bea \label{j_tau_1}
{\bm j}^{(2\omega)}_{21} (\tau_\omega) &=& -\dfrac{e^3 \tau_\omega}{\hbar}\int [d{\bm k}]({\bm E} \times {\bm \Omega})\left[\left(\zeta \Omega_{\rm v} {\bm B}
- \gamma \Omega_{\rm B}  {\bm v}
\right)\cdot  {\bm E} f^\prime~-  \xi  {\bm E} \cdot \left({\bm v}_{\rm m} f^\prime + \epsilon_{\rm m} {\bm v} f^{\prime \prime} \right) \right],
\\ \label{j_tau_2_1}
{\bm j}^{(2\omega)}_{21} (\tau_\omega^2) &=&  -\alpha \dfrac{e^3 \tau_\omega^2}{\hbar} \int [d{\bm k}]({\bm E} \times {\bm \Omega}) \hat L   {\bm v}\cdot  {\bm E} f^\prime,
\\ \label{j_tau_2_2}
{\bm j}^{(2\omega)}_{21} (\tau_\omega, \tau_{2\omega}) &=&  -\dfrac{e^3 \tau_\omega \tau_{2\omega}}{\hbar}\int [d{\bm k}] \left(\eta \Omega_{\rm v}{\bm B} -\xi {\bm v}_{\rm m}\right)({\bm E} \cdot {\bm \nabla}_{\bm k}) {\bm v}\cdot  {\bm E} f^\prime - \dfrac{e^3 \tau_\omega \tau_{2\omega}}{\hbar} \int [d{\bm k}] {\bm v} \Big [ \zeta \dfrac{e}{\hbar} ({\bm E} \cdot {\bm B})( {\bm \Omega} \cdot {\bm \nabla}_{\bm k} ){\bm v}\cdot {\bm E} f^\prime \nn
\\ \label{j_tau_2_3}
&& - \gamma \Omega_{\rm B}  ({\bm E}  \cdot {\bm \nabla}_{\bm k} ) {\bm v}\cdot {\bm E} f^\prime + ({\bm E} \cdot {\bm \nabla}_{\bm k})  \left[ \left(\zeta \Omega_{\rm v} {\bm B}
- \gamma \Omega_{\rm B}  {\bm v}
\right)\cdot  {\bm E} f^\prime - \xi  {\bm E} \cdot \left({\bm v}_{\rm m} f^\prime + \epsilon_{\rm m} {\bm v} f^{\prime \prime} \right) \right]
\Big]~,
\\\label{j_tau_3_1}
{\bm j}^{(2\omega)}_{21} (\tau_\omega, \tau_{2\omega}^2) &=& -\alpha\dfrac{e^3 \tau_\omega \tau_{2\omega}^2}{\hbar} \int [d{\bm k}]{\bm v}\hat L ( {\bm E}  \cdot {\bm \nabla}_{\bm k}) {\bm v} \cdot {\bm E}f^\prime,~~{\bm j}^{(2\omega)}_{21} (\tau_\omega^2, \tau_{2\omega}) = -\alpha\dfrac{e^3 \tau_{2\omega}\tau_\omega^2}{\hbar} \int [d{\bm k}] {\bm v} ({\bm E}  \cdot {\bm \nabla}_{\bm k} ) \hat L {\bm v} \cdot {\bm E} f^\prime.
\eea
\end{widetext}
The NL conductivities extracted from these expressions are presented in the main text. We note that this general formalism for linear $\bm B$-field dependent NL conductivity has been earlier discussed in Refs.~[\onlinecite{Cortijo2016, Morimoto2016c}]. It is straightforward to see that in presence of SIS (TRS broken), all these terms vanish identically. In presence of TRS (broken SIS) however the current $\propto \tau^2$ survives while the contributions $\propto (\tau, \tau^3)$ vanish identically.
%%

%\section{Weyl node separated in momentum space}
%Now, we present our calculation for pair of Weyl nodes that are separated in momentum space. The low energy Hamiltonian we consider for that is given by
%
%\be \label{ham_two_broken}
%H_s ({\bm k}) = s\hbar v_F {\bm \sigma} \cdot ({\bm k} - s {\bm Q}) - sQ_0~.
%\ee
%
%For simplicity we will consider the node separation along the $\hat{\bm z}$ direction only, ${\bm Q} = (0,0,Q)$. The energy dispersion is then given by $\epsilon_s ({\bm k}) = -sQ_{\rm 0} + s_{\rm b} \hbar v_F \sqrt{k_\perp^2 + (k_z - sQ)^{\rm 2}}$, with $k_\perp^{\rm 2} = k_x^{\rm 2} + k_y^{\rm 2}$. The band velocity and geometrical quantities are also modified by the node separation as
%
%\bea
%&& \{ {\bm v},{\bm \Omega}, {\bm m} \} \nn
%\\
%&=& \left\lbrace s_{\rm b} v_F \dfrac{{\bm k} - s {\bm Q}}{|{\bm k} - s {\bm Q}|}, -s e v_F \dfrac{{\bf k} - s {\bm Q}}{2|{\bm k} - s {\bm Q}|^2}, -s s_{\rm b} \dfrac{{\bm k} - s {\bm Q}}{2|{\bm k} - s {\bm Q}|^3} \right\rbrace . \nn
%\\
%&&~~~~~~~~~~~
%\eea
%
%Using these expressions, we calculated the NL magneto-conductivities. We find that it has the same result as the ${\bm Q=0}$ and conclude that inducing momentum separation does not produce yield any different result.

\section{Details of calculation for nonlinear conductivity in WSM}
\label{istrpc_WM}
In this section of Appendix, we present the details of our calculation of NL conductivitity for a pair of Weyl nodes of which one is situated at energy $Q_0$ and the other is situated at energy $-Q_0$. For that we first calculate the NL conductivities of a single Weyl node with band crossing at zero energy given by $H_s ({\bm k}) = s\hbar v_F {\bm \sigma} \cdot {\bm k}$. Then we modify the Fermi energy dependences for individual nodes to include the positional shift in the energy.
%A low energy model Hamiltonian realizing a single node of an isotropic WM is given by~
%
%\be \label{single_ham}
%H_s = s\hbar v_F {\bm \sigma} \cdot {\bm k}
%\ee
%
The velocity for this model Hamiltonian is given by ${\bm v} = s_{\rm b} v_F {\bm k}/k$, while the Berry curvature and OMM are given by~[\onlinecite{Morimoto2016c,Li2021}] ${\bm \Omega} = -s s_{\rm b} {\bm k}/(2k^3)$ and $ {\bm m}= -s e v_F {\bm k}/(2k^2)$.
%
%\be
%\{ {\bm v}, {\bm m}, {\bm \Omega} \} = \left\{ s_{\rm b} v_F \dfrac{{\bm k}}{k}, -s e v_F \dfrac{{\bm k}}{2k^2}, -s s_{\rm b} \dfrac{{\bm k}}{2k^3} \right \} ~,
%\ee
%
The linear Drude conductivities which are diagonal components of the linear conductivity matrix in absence of magnetic field are calculated to be~[\onlinecite{Das2019a}]
\be \label{sgm_aa_istrpc_rslt}
\sigma_{aa} = \dfrac{e^2 \tau \mu^2}{6 \pi^2 \hbar^3 v_F};~~~~a \in (x,y,z).
\ee
Equation~(\ref{sgm_aa_istrpc_rslt}) does not depend on the chirality of the nodes and whether the Fermi levels reside in the conduction band or valence band. In presence of magnetic field the Onsager's reciprocal relations restrict the longitudinal conductivities to be minimum quadratic order in magnetic field which is out of the scope of this paper. However, the off-diagonal components can have minimum linear-${\bm B}$ dependence and are obtained to be~[\onlinecite{Das2021}]
\be \label{sgm_ab_istrpc_rslt}
\sigma_{ab} = - \varepsilon_{abc} B_c \dfrac{s_{\rm b} e^3 v_F}{24 \pi^2 \hbar^3 |\mu|} (\xi \hbar^2 + 4\alpha \tau^2 \mu^2) ~.
\ee
Note that the OMM and Lorentz force cause the ordinary Hall effect.

The magnetic field independent and linear-${\bm B}$ dependent NL conductivities have been summarized in Table~\ref{table_NLC_single}. We obtain the NL anomalous Hall conductivities to be 
\be
\sigma^{\rm NAH}_{abc} = -\varepsilon_{abc}\dfrac{se^3\tau}{12 \pi^2\hbar^2}.
\ee
We note that it is independent of chemical potential and depends on the chirality. So even for Weyl nodes separated in energy its total contribution will vanish. The magnetic field dependent NL conductivities originate from various sources and are written in units of 
\be \label{sigma_NL_S}
\sigma^{\rm S}_{\rm NL} = s s_{\rm b} \frac{\chi_0}{12}~~~ \text{with} ~~~\chi_0 = \frac{e^4 \tau^2 v_F^2}{\pi^2 \hbar^2 |\mu|}.
\ee
The various contributions of magnetic field are also highlighted in the table inside the parenthesis. We note that our results for the NL conductivities are consistent with Ref.~[\onlinecite{Morimoto2016c}], where the effect of Lorentz force effect was ignored. If we ignore the effect of Lorentz force, then for magnetic field along the $\hat {\bm z}$ the expression of $\sigma_{zxx}$ matches with the expression given in Ref.~[\onlinecite{Morimoto2016c}]. %\textcolor{blue}{However, $\sigma_{zzz}$ produces finite result. After considering a more generic treatment of the Berry phase induced terms ($\partial_b \Omega_B\neq 0$) in the conductivity, $\sigma_{zzz}$ vanishes identically.} 

Given the expressions of NL conductivity for Weyl node at zero energy, now we will show how to modify these expressions to obtain results for Weyl nodes separated in energy. We will show this for the Eq.~\eqref{sigma_NL_S} and the modification in linear conductivities can be obtained following similar steps. For scenario-I shown in Fig.~\ref{fig_2}(c) we obtain 
\be
\sigma^{\rm S}_{\rm NL} (+1) = \frac{e^4 \tau^2 v_F^2}{12 \pi^2 \hbar^2 ( |\mu| +Q_0 )};~\sigma^{\rm S}_{\rm NL} (-1) = -\frac{e^4 \tau^2 v_F^2}{12 \pi^2 \hbar^2 ( |\mu| -Q_0 )},
\ee
for scenario-III we obtain 
\be
\sigma^{\rm S}_{\rm NL} (+1) = -\frac{e^4 \tau^2 v_F^2}{12 \pi^2 \hbar^2 ( |\mu| -Q_0 )};~\sigma^{\rm S}_{\rm NL} (-1) = \frac{e^4 \tau^2 v_F^2}{12 \pi^2 \hbar^2 ( |\mu| +Q_0 )},
\ee
and for scenario-II we obtain
\begin{equation} \label{scn-II}
\begin{aligned}
\sigma^{\rm S}_{\rm NL} (+1) = \frac{e^4 \tau^2 v_F^2}{12 \pi^2 \hbar^2 ( \pm |\mu| +Q_0 )}
\\
\sigma^{\rm S}_{\rm NL} (-1) = -\frac{e^4 \tau^2 v_F^2}{12 \pi^2 \hbar^2 ( \pm |\mu| -Q_0 )}
\end{aligned}
\end{equation}
In Eq.~\eqref{scn-II} the $+(-)$ sign in the denominator infront of $|\mu|$ corresponds to $\mu>0 (\mu<0)$ in scenario-II. The total NL conductivities are obtained by adding the contribution from the two nodes. Comparing scenario-I with -III, one can easily identify that
\be
\sigma^{\rm S,I}_{\rm NL} (+1)=\sigma^{\rm S,III}_{\rm NL} (-1), ~\sigma^{\rm S,I}_{\rm NL} (-1)=\sigma^{\rm S,III}_{\rm NL} (+1)~.
\ee
So the NL conductivities will be identical in both these cases after summing over the nodes. Similarly, when we compare scenario-I to -II, it can be checked that 
\bea
\sigma^{\rm S,I}_{\rm NL} (+1)&=&\sigma^{\rm S,II}_{\rm NL} (+1),~~~\sigma^{\rm S,I}_{\rm NL} (-1)=\sigma^{\rm S,II}_{\rm NL} (-1)~\text{for} ~\mu > 0,\nn
\\
\sigma^{\rm S,I}_{\rm NL} (+1)&=&\sigma^{\rm S,II}_{\rm NL} (-1),~~~\sigma^{\rm S,I}_{\rm NL} (-1)=\sigma^{\rm S,II}_{\rm NL} (+1)~ \text{for} ~\mu < 0.~\nn
\\
\eea
So in the above two scenarios also, the NL conductivities turned out to be identical after considering the contributions of all the nodes. In conclusion, the NL conductivities do not depend on the position of the Fermi level.

%%%%%%%%%%%%%%%%%%%%%%%%%%%%%%%%

%%
\bibliography{NLMT_3D.bib}% Produces the bibliography via BibTeX.

\end{document}